\documentstyle[preprint,eqsecnum,aps]{revtex}
\tighten
\begin{document}
\draft
\title{The String Dilaton and a  Least Coupling Principle}
\author{T. Damour}
\address{Institut des Hautes Etudes Scientifiques, 91440 Bures sur Yvette
and,\\
DARC, CNRS-Observatoire de Paris, 92195 Meudon, France}
\author{A.M. Polyakov}
\address{Physics Department, Jadwin Hall, Princeton University,\\ Princeton,
New Jersey 08544, U.S.A.}
\date{\today}
\maketitle
\begin{abstract}
It is pointed out that string-loop modifications of the low-energy
matter couplings of the dilaton may provide a mechanism for fixing the
vacuum expectation value of a massless dilaton in a way which is
naturally compatible with existing experimental data. Under a certain
assumption of universality of the dilaton coupling functions ,
 the cosmological evolution of the graviton-dilaton-matter
system is shown to drive the dilaton towards values where it decouples
from matter (``Least Coupling Principle"). Quantitative estimates are
given of the residual strength, at the present cosmological epoch, of
the coupling to matter of the dilaton. The existence of a weakly coupled
massless dilaton entails a large spectrum of small, but non-zero,
observable deviations from general relativity. In particular, our
results provide a new motivation for trying to improve by several orders
of magnitude the various experimental tests of Einstein's Equivalence
Principle (universality of free fall, constancy of the constants,\dots).
\end{abstract}
\narrowtext

\section{Introduction}

 At present we know only one theory which treats gravity in a way
consistent with quantum mechanics: string theory. In the low energy
limit (low in comparison with the Planck mass) string theory gives
back classical general relativity, with, however, an  important
difference. All versions of string theory predict the existence of a
(four-dimensional) scalar partner of the tensor Einstein
graviton: the dilaton. It may
happen that this scalar field acquires a mass due to some yet unknown
dynamical mechanism. This is the generally adopted view, and if so there
will be no observable macroscopic difference between string gravity and
Einstein gravity. In this paper we will discuss another possibility:
that the dilaton remains massless. This immediately leads to the
dramatic conclusion that all coupling constants and masses of elementary
particles, being dependent on the dilaton scalar field, should be,
generally speaking,  space and time dependent, and influenced by local
circumstances. This conclusion is of course not new and it was
precisely the reason for discarding the possibility that we are
going to discuss.
Indeed, it has been stated that the existence of a massless dilaton
contributing to macroscopic couplings would, at once, entail the
following observable consequences: (i) Jordan-Fierz-Brans-Dicke-type
\cite{JFBD} deviations from Einstein's theory in relativistic
$[O(Gm/c^2r)]$ gravitational effects \cite{SS}; (ii) cosmological
variation of the fine structure constant, and of the other gauge coupling
constants \cite{W}, and (iii) violation of the (weak) equivalence
principle \cite{TV}. As the strength of the coupling of the dilaton
to matter is expected to be comparable to that of the (spin 2) graviton,
and even larger than it in the case of hadrons \cite{TV}, the above
observable consequences seem to be in violent conflict with experiment.
Indeed, present experimental data give upper limits of order: (i)
$10^{-3}$ for a possible fractional admixture of a scalar component to
the relativistic gravitational interaction \cite{Viking}, (ii)
$10^{-15}$yr$^{-1}$ for the fractional variation with time of
the fine-structure constant\footnote{Note that, within the QCD
framework, it does not make sense to speak of the variation of any
strong-interaction coupling constant (the hadron mass-scale adjusting
itself such that $\alpha_{\rm strong} \simeq 1$).} \cite{SV90},
(iii) $10^{-11}$
---~$10^{-12}$ on the universality of free fall (weak equivalence
principle) \cite{RKD,BP,A,LLR}\,\footnote{The most recent analysis
of Lunar Laser Ranging data \cite{LLR} finds that the
fractional difference in gravitational
acceleration toward the Sun between the
(silica-dominated) Moon and the (iron-dominated) Earth
 is $(-2.7 \pm 6.2)
\times 10^{-13}$.}. [See \cite{W92,D93} for reviews of the
comparison between gravitational theories and experiments].

In this paper, we point out that non-perturbative string loop effects
(associated with worldsheets of arbitrary genus in intermediate string
states) can naturally reconcile the existence of a massless dilaton
with existing experimental data if they exhibit the same kind of
universality as the tree level dilaton couplings.
By studying the cosmological evolution
of general graviton-dilaton-matter systems we show that the dilaton
is cosmologically attracted toward values where it decouples from
matter, a situation which we call the ``Least Coupling Principle''.
Roughly speaking, the origin of the attraction is the following.
Masses of different particles depend on the dilaton, while the source
for the dilaton is the gradient of these masses. It is therefore
not surprising to have a fixed point where the gradient of the masses
is zero.
 [With some important differences discussed below, this mechanism
is similar to the generic attractor mechanism of metrically-coupled
tensor-scalar theories discussed in Refs. \cite{DN}].
This cosmological attraction is so efficient
that the presently existing experimental limits do not place any
significant constraints on the physical existence of a massless dilaton.
Most importantly, we give quantitative estimates for the level of
residual deviation from Einstein's theory expected at the present
cosmological epoch, notably for the violation of the equivalence
principle.

\section{The graviton-dilaton-matter system}

At the tree level in the string loop expansion (spherical topology
for intermediate worldsheets) the effective action describing the
massless modes (here considered directly in four dimensions) has the
general form \cite{FT,CFMP,CKP}
\begin{eqnarray}
 S_{\rm tree} = \int d^4x &&\sqrt{\hat g} e^{-2\Phi} \{ (\alpha')^{-1}
[\widehat R +4\,\widehat{\Box} \Phi - 4 (\widehat\nabla \Phi)^2]\nonumber \\
&& -{k\over 4} \widehat F^a_{\mu\nu} \widehat F^{a\mu\nu}
    - \overline{\widehat \psi} \widehat D \widehat \psi + \cdots \nonumber\\
&& + \sum_{n\geq 1} O[( \alpha' \partial^2)^n] \}\ . \label{eq:2.1}
\end{eqnarray}
Here, $\hat g_{\mu\nu}$ (often denoted $G_{\mu\nu}$) denotes the metric
appearing in the $\sigma$-model formulation of string theory and is
used for defining all the covariant constructs entering Eq.~(1)
$[\widehat \nabla, (\widehat F)^2, \widehat D, \cdots ]$; $\Phi$ denotes
the dilaton; a summation over the various possible gauge fields
$[F^a_{\mu\nu}= \partial_\mu A^a_\nu -\partial_\nu A^a_\mu + f^{abc}
A^b_\mu A^c_\nu ]$ and fermions $[\widehat D = \widehat\gamma^\mu
(\widehat\nabla_\mu + A^a_\mu t^a)]$ is understood; the ellipsis stand
in particular for the ill-understood remaining scalar sector of the
theory [Higgs fields and their Yukawa couplings, and possibly other
gauge-neutral scalar (moduli), or pseudo-scalar (axion,\dots), fields];
and the last term symbolically denotes the infinite series of
higher-derivative terms representing the low-energy effects of all the
massive string modes on which one has to integrate to get the effective
action for the massless modes.

The remarkable feature that, when formulating the action in terms of
the ``string frame" metric $\hat g_{\mu\nu}$, the dilaton couples, at
the string tree level, in a universal, multiplicative manner to all the
other fields derives from the fact that $g_s \equiv \exp (\Phi)$ plays
the role of the string coupling constant. In the $\sigma$-model
formulation, this is easily seen to follow from applying the
Gauss-Bonnet theorem $[(4\pi)^{-1} \int d^2 \xi \sqrt h R^{(2)} (h) =
\chi = 2 (1-n)$; $n$= number of handles] to the Fradkin-Tseytlin
$\sigma$-model
dilaton term, $S_{\rm dil} =(4\pi)^{-1} \int d^2 \xi \sqrt h \Phi (X)
R^{(2)}$. In a constant (or slowly varying) dilaton background, the
genus-$n$ string-loop contribution to any string transition amplitude
contains the factor $\exp (-S_{\rm dil}) =\exp (2(n-1)\Phi) = g^{2(n-1)}_s$.
Therefore, when taking into account the full string loop expansion,
the effective action for the massless modes will take the general form
\begin{eqnarray}
 S = \int d^4 x\, \sqrt{\hat g}&& \left\{ {B_g(\Phi)\over \alpha '}\,
 \widehat R + {B_\Phi(\Phi)\over \alpha '}\, [ 4 \widehat\Box \Phi -
  4 (\widehat\nabla \Phi)^2] \right. \nonumber\\
 && \left.  - B_F (\Phi) {k\over 4} \widehat F^2 - B_\psi (\Phi)
    \overline{\widehat\psi} \widehat D \widehat\psi + \cdots \right\}\ .
  \label{eq:2.2}
\end{eqnarray}
At this stage of development of string theory, one does not know how
to control the structure of the various dilaton coupling functions
$B_i (\Phi)$ $(i=g,\Phi,F,\psi,...)$ beyond the fact that in the limit $\Phi
\to -\infty$ $(g_s\to 0)$ they should admit an expansion in powers of
$g_s^2 =\exp (+2\Phi)$ of the form,
\begin{equation}
 B_i (\Phi) = e^{-2\Phi} +c^{(i)}_0 + c^{(i)}_1 e^{2\Phi} + c^{(i)}_2
  e^{4\Phi} + \cdots \label{eq:2.3}
\end{equation}
[Note that we have in mind the low-energy regime , with broken
supersymmetry, for which there are no  {\it a priori} obstacles
to having couplings of the type (\ref{eq:2.3}) with
$c^{(i)}_n \neq 0$.]

Concerning the low-energy effects of all the massive string modes, we shall
assume for simplicity that, like at tree level, Eq.~(\ref{eq:2.1}), they
are equivalent to introducing a cut-off at a \underbar{$\Phi$-independent}
string mass scale $\widehat\Lambda_s\sim (\alpha ')^{-1/2}$, when
measuring distances by means of the \underbar{string-frame} metric $\hat
g_{\mu\nu}$.

It is convenient to transform the action (\ref{eq:2.2}) by introducing
several $\Phi$-dependent rescalings.  One can put both the gravity and
the fermion sectors into a standard form by:  (i) introducing the
``Einstein metric",
\begin{equation}
 g_{\mu\nu} \equiv C\ B_g (\Phi) \hat g_{\mu\nu} \label{eq:2.4}
\end{equation}
(with some numerical constant $C$~\footnote{We shall choose $C$ such that
the string units and the Einstein units coincide at the present cosmological
epoch: $C B_g(\Phi_0) = 1$.}), (ii) replacing the original dilaton
field $\Phi$ by the variable\footnote{When $B_g = B_{\Phi}$ the quantity
under the square root in Eq.(\ref{eq:2.5}) is  positive definite.}
\begin{equation}
\varphi \equiv \int d\Phi \left[ {3\over 4}\left({B'_g\over B_g}\right)^2
 + 2 {B'_\Phi\over B_g} + 2 {B_\Phi\over B_g}  \right]^{1/2}
  \label{eq:2.5}
\end{equation}
(where a prime denote $d/d\Phi$),
and (iii) rescaling the Dirac fields
\begin{equation}
 \psi \equiv C^{-3/4} B_g^{-3/4} B^{1/2}_\psi \widehat\psi\ .\label{eq:2.6}
\end{equation}
The transformed action can be decomposed into a gravity sector
$(g_{\mu\nu},\varphi)$ and a matter one $(\psi, A, \cdots)$
\begin{mathletters}
\label{eq:2.7}
\begin{equation}
 S [g,\varphi, \psi, A,\cdots] = S_{g,\varphi} + S_m \ , \label{eq:2.7a}
\end{equation}
\begin{eqnarray}
 S_{g,\varphi} &=& \int d^4x\, \sqrt g \left\{ {1\over 4q} R
    - {1\over 2q} (\nabla \varphi)^2 \right\}\ , \label{eq:2.7b} \\
 S_m &=& \int d^4 x\, \sqrt g \left\{ -\overline\psi\, D\, \psi -
 {k\over 4}\, B_F(\varphi) F^2 + \cdots \right\}\ . \label{eq:2.7c}
\end{eqnarray}
\end{mathletters}
Here, $q\equiv 4\pi \overline G \equiv {1\over 4} C \alpha '$ ($\overline
G$ denoting a \underbar{bare} gravitational coupling constant),
$B_F(\varphi) \equiv B_F [\Phi (\varphi)]$ and the ellipsis stand for
the (more complicated) Higgs sector. One should note that the string
cut-off mass scale acquires a dependence upon the dilaton in Einstein
units:
\begin{equation}
 \Lambda_s (\varphi) \equiv C^{-1/2} B_g^{-1/2} (\varphi) \widehat\Lambda_s
 \ . \label{eq:2.8}
\end{equation}

Essential to the following will be the dilaton dependence of the matter
Lagrangian. One does not know at present how to relate string models to
the observed particle spectrum. The basic clue that we shall follow
is the dilaton dependence of the gauge coupling constants: $g^{-2} =k\,
B_F(\varphi)$ from (\ref{eq:2.7c}). To connect the (bare) effective
action (\ref{eq:2.7}) (integrated over the massive string modes) to the
low-energy world, one still needs to take into account the quantum
effects of the light modes between the string scale $\Lambda_s(\varphi)$
and some observational scale. In the case of an asymptotically free
theory the ratio of the IR confinment mass scale $\Lambda_{\rm conf}$ to
the cut-off scale, is, at the one-loop level, exponentially related to
the inverse of the gauge coupling constant appearing in the bare action:
\begin{equation}
 \Lambda_{\rm conf} \sim \Lambda_s \exp (-8\pi^2 b^{-1} g^{-2}) =
 C^{-1/2} B_g^{-1/2} (\varphi) \exp [-8\pi^2 b^{-1}k\,B_F(\varphi)]
 \widehat\Lambda_s \ , \label{eq:2.9}
\end{equation}
where the one-loop coefficient $b$ depends upon the considered gauge
field as well as the matter content.
The mass of hadrons is, for the most part, generated  by QCD-effects
and is simply proportional to $\Lambda_{QCD}$ ( with some pure number as
proportionality constant). The dilaton dependence of the QCD part of the mass
of hadrons is therefore given by (\ref{eq:2.9}) with $b = b_3$ and
$B_F = B_3$ being the appropriate QCD quantities. However, the lepton
masses, and the small quark contributions to the mass of hadrons, are
not related to $\Lambda_{QCD}$ ( at least in any known way). Their
dilaton dependence is defined by specific mechanisms of spontaneous
symmetry breakdown (and compactification) which depend on particular
string models and are not well established at present. Let us note
that in technicolor-type models, as well as in no-scale
supergravity ones, all the particle mass scales
are related to the fundamental cut-off
 scale by formulas of the type (\ref{eq:2.9}).

As a minimal ansatz, we can assume that the mass
 (in Einstein units) of any type
of particle, labelled $A$, depends in a non-trivial way on the VEV
of the dilaton through some of the functions $B_i$ appearing in
(\ref{eq:2.2}):
\begin{equation}
 m_A(\varphi) = m_A [B_g (\varphi) ,\  B_F(\varphi),\cdots ] \ .
 \label{eq:2.10}
\end{equation}
The essential new feature allowed by nonperturbative string-loop effects
(i.e. arbitrary functions $B_i(\Phi)$, Eq.~(\ref{eq:2.3})) is the
possibility for the function $m_A(\varphi)$ to admit a minimum for some
finite value of $\varphi$. Assuming this, we shall see below that the
cosmological evolution naturally attracts $\varphi$ to such a minimum.
However, if the various coupling functions $B_i(\varphi)$
differ from each other the minima of $m_A(\varphi)$ depend , in
general, on the type of particle considered. It will be seen below
that this weakens the attraction effect of the cosmological
expansion,and, more importantly, leaves room for violations of
the equivalence principle at a probably unacceptable level
(see the footnote following Eq.(\ref{eq:6.13})). This suggests
to concentrate on the case where string-loop effects preserve the
universal multiplicative coupling present at tree-level,
Eq.(\ref{eq:2.1}), i.e. the case where
all the dilaton coupling functions coincide: $B_i(\varphi) =
B(\varphi)$ for $i=F,g,\cdots$. In this ``universal $B(\varphi)$''
case, the extrema of the function $m_A(\varphi) = m_A[B(\varphi)]$
will (generically) coincide with the extrema of the function
$B(\varphi)$. As discussed below, this assumption leads very
naturally (without fine-tuning, or the need to inject small
parameters) to a situation where the present deviations from
general relativity are so small as to have escaped detection.
When we shall need in the following to estimate quantitatively the
dependence of particle masses on $\varphi$, we shall assume
that the mass of any particle $A$ is of the form suggested
by Eq.(\ref{eq:2.9}):
\begin{equation}
 m_A(\varphi) = \mu_A B^{-1/2} (\varphi) \exp [-8\pi^2 \nu_A B(\varphi)]
 \widehat\Lambda_s \ , \label{eq:2.11}
\end{equation}
with $\mu_A$ and $\nu_A$ pure numbers of order unity. We believe
that our main qualitative conclusions do not depend strongly on
the specific form of the assumption (\ref{eq:2.11}).

 For the
quantitative estimates below we need to choose some specific value of
the string unification scale $\widehat\Lambda_s \propto \alpha'^{-1/2}$.
The theoretical value $\widehat\Lambda_s = e^{(1-\gamma)/2} 3^{-3/4}
g_s M_{\rm Planck} /4\pi \simeq g_s \times 5.27 \times 10^{17}$GeV  has
been suggested \cite{K88}. Here $g_s$ denotes the common (modulo possible
Kac-Moody level factors of order unity) value of the gauge coupling
constants at the string scale. To fix ideas,  we shall take
$\widehat\Lambda_s =3\times 10^{17}$GeV.

\section{Classical cosmology with a dilaton}

 The gravitational field equations derived from Eqs.~(\ref{eq:2.7}) read
\begin{mathletters}
\label{eq:3.1}
\begin{eqnarray}
 R_{\mu\nu} &=& 2\partial_\mu \varphi\, \partial_\nu\varphi
 + 2q \left( T_{\mu\nu} - {1\over 2} T g_{\mu\nu} \right)\ ,\label{eq:3.1a}\\
 \Box \varphi &=& - q\sigma \ , \label{eq:3.1b}
\end{eqnarray}
\end{mathletters}
where the source terms, defined by $T^{\mu\nu}\equiv 2g^{-1/2} \delta
S_m/\delta g_{\mu\nu}$, $\sigma \equiv g^{-1/2} \delta S_m /\delta
\varphi$, are related by the energy balance equation: $\nabla_\nu
T^{\mu\nu} = \sigma \nabla^\mu \varphi$.

 In the case of a Friedmann cosmological model, $ds^2=-dt^2 +a^2(t)
d\ell^2$ with $d\ell^2 =(1-Kr^2)^{-1}dr^2 +r^2 (d\theta^2 +\sin^2\theta
d\varphi^2)$, $K=0,$ $+1$ or $-1$, the field equations give
$(T^\mu_\nu ={\rm diag} (-\rho ,P,P,P)$; $H\equiv \dot a/a$, the
overdot denoting $d/dt$)
\begin{mathletters}
\label{eq:3.2}
\begin{equation}
 -3 {\ddot a\over a} = q(\rho + 3P) + 2\dot \varphi^2\ , \label{eq:3.2a}
\end{equation}
\begin{equation}
 3H^2 + 3{K\over a^2} = 2 q \rho + \dot\varphi^2\ , \label{eq:3.2b}
\end{equation}
\begin{equation}
 \ddot\varphi + 3H\,\dot\varphi = q\,\sigma\ . \label{eq:3.2c}
\end{equation}
\end{mathletters}
 In the following, we concentrate on the spatially flat case ($K=0$).
Following Ref.~\cite{DN}, we can combine Eqs.~(\ref{eq:3.2}) to
write a simple equation for the cosmological evolution of the dilaton
with respect to the logarithm of the cosmological scale factor:
$p\equiv \ln (a) + {\rm const.}$ (not to be confused with the pressure $P$).
Denoting $d/dp$ by a prime, one gets $(K=0)$
\begin{equation}
 {2\over 3-\varphi'^{2}} \varphi'' + (1- \lambda) \varphi'
    = {\sigma\over\rho} \ , \label{eq:3.3}
\end{equation}
where $\lambda \equiv P/\rho$.

Except during phase transitions, the material content of the universe
can be classically described as a superposition of several (weakly
interacting) gases labelled by $A$, i.e. by an action of the form
\begin{equation}
S_m [g,\varphi, x_A] = - \sum_A \int m_A [\varphi (x_A)] [-g_{\mu\nu}
(x^\lambda_A) dx^\mu_A dx^\nu_A]^{1/2} \label{eq:3.4}
\end{equation}
(the massless particles being obtained by taking the limit $m_A\to 0$
with $m_Au^\mu_A \equiv m_Adx^\mu_A /ds_A$ fixed). In Eq.~(\ref{eq:3.4})
the summation over $A$ includes a sum over the statistical distribution
of the $A$-type particles. The gravitational source terms corresponding
to Eq.~(\ref{eq:3.4}) read
\begin{mathletters}
\label{eq:3.5}
\begin{eqnarray}
 T^{\mu\nu} (x) &=& {1\over \sqrt{g(x)}} \sum_A \int ds_A m_A [\varphi (x_A)]
 u^\mu_A u^\nu_A \delta^{(4)} (x-x_A)\ , \label{eq:3.5a} \\
 \sigma (x) &=& -{1\over \sqrt{g(x)}}\sum_A \int ds_A\alpha_A[\varphi (x_A)]
 m_A [\varphi (x_A)] \delta^{(4)} (x-x_A)\nonumber\\
 &=& \sum_A \alpha_A [(\varphi (x)] \ T_A(x) \ , \label{eq:3.5b}
\end{eqnarray}
\end{mathletters}
where
\begin{equation}
 \alpha_A (\varphi) \equiv {\partial\ln\, m_A(\varphi)\over \partial\varphi}
  \label{eq:3.6}
\end{equation}
measures the strength of the coupling of the dilaton to the $A$-type
particles. In the second Eq.~(\ref{eq:3.5b}) $T_A = -\rho_A +3P_A$ denotes
the trace of the $A$-type contribution to the total $T^{\mu\nu}
=\Sigma_A T^{\mu\nu}_A$. It is easy to see that when the different
$A$-gases are non interacting their corresponding sources satisfy the
separate energy balance equations: $\nabla_\nu T^{\mu\nu}_A = \sigma_A
\nabla^\mu\varphi =  \alpha_A T_A \nabla^\mu \varphi$.

In the string context, it is natural to assume that the string scale
$\Lambda_s$ subsumes both what is usually meant by ``Planck scale" and
``GUT scale", leaving essentially no room for a quasi-classical
inflationary era. We leave to future work a discussion of primordial
stringy cosmology, and content ourselves by describing the evolution of
dilatonic cosmologies through a radiation-dominated era, followed by a
matter-dominated one.

\section{Evolution of the dilaton during the radiation-dominated era.}

 During a radiation-dominated era (universe dominated by
ultra-relativistic gases) the gravitational source terms are
approximately given by
\begin{mathletters}
\begin{equation}
 \rho \simeq 3P \simeq g_\ast (T) {\pi^2\over 30} T^4\ , \label{eq:4.1a}
\end{equation}
\begin{equation}
 \sigma \simeq 0 \ , \label{eq:4.1b}
\end{equation}
\end{mathletters}
where $g_\ast (T) = \sum_{\rm Bose} g^B_A (T_A/T)^4 + (7/8) \sum_{\rm
Fermi} g^F_A (T_A/T)^4$  is the effective number of relativistic degrees
of freedom in the cosmic soup at temperature $T$. [ The sum defining
$g_\ast (T)$ is taken only over particles with mass $m_A \ll T$; because
of possible previous decouplings the corresponding relativistic gases
may not all have the temperature $T$, e.g. $ T_\nu = (4/11)^{1/3} T_\gamma$
 below 1 MeV ].
 Eq.~(\ref{eq:4.1b})
suggests that the dilaton does not evolve during the radiation era. More
precisely, Eq.~(\ref{eq:3.3}) with $\lambda \simeq 1/3$ shows that
$\varphi (p)$ behaves as a particle, with velocity-dependent mass,
submitted to a constant friction. In a few $p$-time units, $\varphi (p)$
will exponentially come to rest. [see Ref.~\cite{DN} for the exact
solution of the damped evolution of $\varphi (p)$ when $\sigma/\rho$ is
negligible]. However, something interesting happens each time the
universe cools down to a temperature $T\sim m_A$ defining the threshold
for the participation of the species $A$ to the relativistic soup. When
$T\sim m_A$, the term  on the right-hand side of the $p$-time evolution
of $\varphi$ is well approximated by
\begin{equation}
{\sigma_A\over \rho_{\rm tot}} = -{15\over \pi^4}\, {g_A\over g_\ast (T)}
\tau_\pm (z_A) \alpha_A (\varphi) \ , \label{eq:4.2}
\end{equation}
where $z_A \equiv m_A/T$ and
\begin{equation}
 \tau_\pm (z) \equiv z^2 \int^\infty_z dx\
    {(x^2-z^2)^{1/2}\over e^x \pm 1} \ ,   \label{eq:4.3}
\end{equation}
where the upper (lower) sign corresponds to $A$ being a fermion (boson).
In the approximation (justified by the results to be discussed) where
the dilaton contributions to the Einstein equations (\ref{eq:3.2a}),
(\ref{eq:3.2b}) are negligible one has $T\propto a^{-1}$, and therefore
$p=\ln z_A$ (with an adapted choice of origin for $p$). Then, as a
function of $p$, the (everywhere positive) function $\tau_\pm$ is
proportional to $\exp (+2p)$ when $p\to -\infty$, rises up to a
maximum $\simeq 1.16$ when $z_A\simeq 0.87$, and falls quickly to zero
as $\exp [{5\over 2} p-\exp (p)]$ when $p\to +\infty$. [ This maximum
occurs because for $T \gg m_A$ the (ultra-relativistic) particles do
not contribute to $\sigma$,  while for  $T \ll m_A$ there are
exponentially few particles ].
 Remembering the
definition (\ref{eq:3.6}) of $\alpha_A(\varphi)$, it is easy to see that
if the function $m_A(\varphi)$ has a minimum, say $\varphi^A_m$, and if
the initial value of $\varphi$, say $\varphi_-^A \equiv \varphi
(p=-\infty) = \varphi (T\gg m_A)$, is sufficiently near $\varphi^A_m$,
Eq.~(\ref{eq:3.3}) will describe a damped, transient nonlinear
attraction of $\varphi (p)$ around $\varphi^A_m$. [Note that, from the
above discussion, the initial velocity is zero to an exponential
accuracy $\sim \exp (p_{A'} -p_A)$]. The existence of such an attraction
mechanism by mass thresholds during the radiation era was noticed in
Ref.~\cite{DN} in a related context (generalized
Jordan-Fierz-Brans-Dicke theories characterized by a universal,
$A$-independent coupling function $\alpha_A(\varphi) = \alpha (\varphi)
= \partial a(\varphi) /\partial\varphi$). In the context of
Ref.~\cite{DN}, it seemed natural to assume that the curvature of
the function $\ln m(\varphi) = a(\varphi) +\ln m_0$ near its minimum
was of order unity. This rendered the presently discussed attraction
mechanism very ineffective. An important new feature of the present,
dilatonic, context is that the curvature of $\ln m_A (\varphi)$ near
its minimum is expected to be large compared to one. This follows from
the expected exponential dependence on $B_F(\varphi)$ of the mass scales
of the low-energy particle spectrum, Eq.~(\ref{eq:2.9}). To fix ideas
and be able to make some quantitative estimates, we shall take the form
(\ref{eq:2.11}) with $\mu_A =1$. This yields
\begin{equation}
\alpha_A (\varphi) = - \left[ \ln {\widehat \Lambda_s\over m_A} +{1\over 2}
  \right] {\partial \ln B(\varphi)\over \partial\varphi} = + \ln
  {\widehat\Lambda'_s\over m_A}\,
  {\partial \ln B^{-1}(\varphi)\over \partial\varphi}\ , \label{eq:4.4}
\end{equation}
where $\widehat\Lambda'_s = e^{1/2}\widehat\Lambda_s \simeq 5\times 10^{17}$%
GeV.

We see that a minimum $\varphi_m$ of $m_A(\varphi)$ corresponds to a
maximum of $B(\varphi)$ (or a minimum of $B^{-1}(\varphi)$). Let us
denote by $\kappa$ the curvature of the function $\ln B^{-1}(\varphi)$
near its minimum $\varphi_m$. In the parabolic approximation
\begin{equation}
  \ln B^{-1} (\varphi) \simeq \ln B^{-1} (\varphi_m) + {1\over 2} \kappa
 (\varphi -\varphi_m)^2\ , \label{eq:4.5}
\end{equation}
one gets
\begin{mathletters}
\label{eq:4.6}
\begin{equation}
 \alpha_A (\varphi) = \beta_A (\varphi - \varphi_m)\ , \label{eq:4.6a}
\end{equation}
\begin{equation}
 \beta_A = \kappa\, \ln (\widehat \Lambda'_s/m_A) = \kappa [40.75 - \ln
  (m_A/1\rm {GeV})]\ . \label{eq:4.6b}
\end{equation}
\end{mathletters}
 Inserting Eq.~(\ref{eq:4.6a}) into Eq.~(\ref{eq:4.2}) and then into
Eq.~(\ref{eq:3.3}) (written in the approximation $\lambda \simeq 1/3$)
yields
\begin{equation}
 \varphi'' (p) + \varphi' (p) = s_A (p) [\varphi (p) - \varphi_m]\ ,
 \label{eq:4.7}
\end{equation}
with
\begin{equation}
 s_A (p) = - {45\over 2\pi^4}\, \beta_A {g_A\over g_{\ast(T)}}\
 \tau_\pm (e^p)\ . \label{eq:4.8}
\end{equation}
Within a good approximation one can replace the temperature-dependent
quantity $g_A/g_*(T)$ by its initial value, say $f^{\rm in}_A \equiv
g_A/g_*^{\rm in}$, in which $g_*^{\rm in} \equiv g_* (T\gg m_A)$
 contains the contribution
$7g_A/8$ (or $g_A$) if $A$ is a fermion (or boson). Eq.~(\ref{eq:4.7})
describes a damped motion submitted to a transient harmonic force tending
to attract $\varphi$ toward $\varphi_m$. The final outcome of this motion
is to leave (when $p=+\infty$) $\varphi$ nearer to $\varphi_m$ than it
was when it started at rest at $p=-\infty$. We define the attracting
factor of the $A$-th mass threshold as $m_\pm (b_A)\equiv (\varphi
(+\infty)-\varphi^A_m)/(\varphi(-\infty)-\varphi^A_m)$, where
the suffix $\pm$ in the left-hand side
corresponds to the fermion/boson case and where $b_A\equiv \beta_A
f^{\rm in}_A\equiv \beta_A g_A/g^{\rm in}_*$. There are two quite
different regimes in this mass-threshold attraction mechanism: when
$b_A\ll 1$ ($b_A <0.5$ sufficing), $\varphi (p)$ moves monotonically
toward $\varphi_m$ by a small amount given by integrating over $p$ the
force term on the right-hand side of (\ref{eq:4.7}) evaluated at the
original position of $\varphi$ (``kick" approximation). The result is
(see Ref.~\cite{DN})
\begin{equation}
 m_\pm (b_A) = 1 - {1\over 2} b_A {7/8\choose 1} + O(b^2_A)\ ,
\label{eq:4.9}
\end{equation}
where the upper (lower) coefficient corresponds to the fermion (boson)
case, respectively. In this first case the attracting power of the
$A$-threshold is rather weak (hence the conclusion of Ref.~\cite{DN}
that the total radiation era attraction is rather ineffective in the
case of usual tensor-scalar theories with $\beta_A=O(1)$ and $\Sigma_A
f_A \sim \ln (100/10) \simeq 2.3)$. By contrast, in the present,
dilatonic context one expects $\kappa \sim 1$, $\beta_A \sim 40$ and
therefore $b_A \gg 1$ for many mass thresholds (the most efficient mass
thresholds being the latest in the radiation era which tend to have the
largest $f^{\rm in}_A$'s: notably the $e^+e^-$ threshold with $f^{\rm
in}_e = 4/10.75\simeq 0.372$). When $b_A\gg1$ ($b_A >2$ sufficing in
practice) one can analytically solve Eq.~(\ref{eq:4.7}) by a WKB-type
approach. [With some subtleties  compared to the usual WKB approximation
as the matching between the damped and
oscillating regions must be done via Bessel functions instead of the
usual Airy ones]. Qualitatively the motion of $\varphi (p)$ begins by a
slow roll toward $\varphi_m$, continues by WKB oscillations around
$\varphi_m$, and terminates as a damped inertial motion. The final
analytical results for the attraction factor reads $(b_A \gg 1)$
\begin{mathletters}
\label{eq:4.10}
\begin{equation}
 m_\pm (b_A) = (C_\pm b_A)^{-1/4} \cos \theta^A_{\pm}\ , \label{eq:4.10a}
\end{equation}
with $C_+ = 15/8$, $C_- =15/4$ and
\begin{equation}
 \theta^A_\pm = \int^{+\infty}_{-\infty} [-s_A (p)]^{1/2} dp - {\pi\over
4} = b^{1/2}_A I_\pm - {\pi\over 4}\ , \label{eq:4.10b}
\end{equation}
\end{mathletters}
with $I_+ \simeq 1.2743$, $I_- \simeq 1.4029$. Note that when $b_A\to
\infty$, $|m_\pm (b_A)|$ tends to zero as $O(b_A^{-1/4})$. Fig.~1
represents the two functions $m_\pm (b)$, obtained by numerically
integrating Eq.~(\ref{eq:4.7}).

One must take into consideration the fact that mass thresholds can occur
only for particles whose masses are smaller than the critical
temperature of the phase transition through which they acquired a mass
(e.g. the pions are the only hadrons to take into account). The Higgs
threshold is to be considered as part of the electroweak phase transition,
and the strange quark threshold overlaps with the quark-hadron phase
transition. This leaves nine, clearly present, mass thresholds
associated (in decreasing temperature scale) with the top quark
$(f^{\rm in}_t =12/106.75$, $\beta_t \simeq 35.74\kappa)$, the $Z^0$
$(f^{\rm in}_Z =3/95.25$, $\beta_Z \simeq 36.24\kappa)$, the $W^\pm$
$(f^{\rm in}_W =6/92.25$, $\beta_W \simeq 36.37\kappa)$, the bottom
quark $(f^{\rm in}_b =12/86.25$, $\beta_b \simeq 39.14\kappa)$, the tau
$(f^{\rm in}_\tau =4/75.75$, $\beta_\tau \simeq 40.17\kappa)$, the charmed
quark $(f^{\rm in}_c =12/72.25$, $\beta_c \simeq 40.35\kappa)$, the pions
$(f^{\rm in}_\pi =3/17.25$, $\beta_\pi \simeq 42.74\kappa)$, the muon
$(f^{\rm in}_\mu =4/14.25$, $\beta_\mu \simeq 43.00\kappa)$, and the
electron $(f^{\rm in}_e =4/10.75$, $\beta_e \simeq 48.33\kappa)$.
The quoted values of $f^{\rm in}_A$ and $\beta_A$ show that $b_A\equiv
\beta_A f^{\rm in}_A$ is typically a few times $\kappa$ (with extreme
values $1.14\kappa$ and $17.98\kappa$ for the $Z$ and $e$ respectively).
A look at Fig.~1 shows immediately that if $\kappa$ is of order unity,
each mass threshold will a be rather efficient attractor. The compound
effect of all those attractors is discussed below.

Besides mass thresholds, phase transitions provide another possible
attractor mechanism for the dilaton during the radiation-dominated era.
During a phase transition the vacuum energy density $V$ changes from
some positive value, say $V^{\rm in} =g_{\rm vac} (\pi^2/30) T^4_c$,
when $T>T_c$ to a comparatively negligible value when $T<T_c$. For
instance, in the case of the QCD (quark-hadron) phase transition one
has $T_c \simeq 200$~MeV and $g_{\rm vac}=34/3$ (in a simple model\cite{B}
describing the unconfined phase as a relativistic gas of gluons and
$u$ and $d$ quarks ---~besides $\gamma ,e, \nu$ and $\mu$~--- and the
confined phase as a relativistic gas of pions). Besides its dependence
on the temperature the vacuum energy density is also a function of the
dilaton. Therefore the vacuum term in the matter action, $S_{\rm vac} =
- \sqrt g V (\varphi ,T)$, will generate a corresponding source term
$\sigma_{\rm vac}= -\partial V/\partial \varphi$ in the right-hand sides
of the dilaton evolution equations (\ref{eq:3.1b}), (\ref{eq:3.2c}) or
(\ref{eq:3.3}). In the simple model of the QCD phase transition just
described, one can estimate the source term $\sigma_{\rm vac}$ by
assuming that the dilaton dependence of $V$ is essentially contained
in the $\varphi$-dependence of the critical temperature $T_c$. In turn,
the latter dependence is obtained from $T_c\sim \Lambda_{QCD}$ with
$\Lambda_{QCD} (\varphi)$ given by Eq.~(\ref{eq:2.9}) with the
appropriate one-loop coefficient. This shows that $\varphi$ will be
attracted toward a minimum of $\Lambda_{QCD} (\varphi)$. More precisely,
if we assume, to fix ideas, that $B_g(\varphi) =B_F(\varphi)$ in
Eq.~(\ref{eq:2.9}) and that $\varphi$ is near the maximum $\varphi_m$
of $B(\varphi)$, one gets, in the parabolic approximation (\ref{eq:4.5})
(setting as above $\lambda =P_{\rm tot}/\rho_{\rm tot}$ and
$p = \ln (T_c/T)$ )
\begin{equation}
 \varphi'' (p) + {3\over 2} [ 1 - \lambda (p)] \varphi '(p) =
 s_{\rm vac} (p) [\varphi (p) - \varphi_m]\ , \label{eq:4.11}
\end{equation}
where $s_{\rm vac} (p) \simeq -6\beta_{\rm vac} f_{\rm vac} \exp (4p)$
when $p \to -\infty$, with $\beta_{\rm vac} =\kappa \ln (\widehat\Lambda'_s
/T_c)$ and $f_{\rm vac} =g_{\rm vac} /g_* (T>T_c)$. After the phase
transition, when $p\to +\infty$, one expects $s_{\rm vac}(p)$ to fall
quickly to zero as $\exp (-aT_c/T) =\exp (-a\exp (p))$ with $a$ of
order unity. In the limit where $b_{\rm vac}\equiv \beta_{\rm vac}
f_{\rm vac}$ is large enough to make $\varphi$ oscillate around $\varphi_m$,
one can solve Eq.~(\ref{eq:4.11}) by a WKB-type approach. The final
result for the attraction factor due to a phase transition, $p(b_{\rm
vac}, \cdots) \equiv (\varphi (+\infty) -\varphi_m)/(\varphi (-\infty)
-\varphi_m)$, reads
\begin{equation}
p(b_{\rm vac}, \cdots) = 2^{3/2} \pi^{-1} \Gamma (5/4) a^{1/2} (6 b_{\rm
vac})^{-1/8} \exp (-I) \cos\theta \ , \label{eq:4.12}
\end{equation}
where $I = (1/4) \int^{+\infty}_{-\infty} [1-3\lambda (p)] dp$, and
where the angle $\theta$ depends on the two functions $\lambda (p)$ and
$s_{\rm vac} (p)$. [In the approximation $\lambda (p) = 1/3$, one finds
$\theta = \int^{-\infty}_{+\infty} [-s_{\rm vac} (p)]^{1/2} dp -\pi/8]$.
In the case of the QCD phase transition, one has $\beta_{\rm vac}\simeq
42.36\kappa$ and $f_{\rm vac} =(34/3)/51.25 \simeq 0.2211$. If $\kappa$
is of order unity, $b_{\rm vac} \simeq 9.37 \kappa$ is probably large
enough to render valid the WKB result (\ref{eq:4.12}). This yields an
attraction factor $p_{QCD} \simeq 0.49 a^{1/2} \kappa^{-1/8} \cos\theta$.
 In the
case of the electroweak phase transition, rough estimates give $b_{\rm
vac} \simeq (\lambda /4)\kappa$ where $\lambda$ denotes the quartic
self-coupling of the Higgs. Its seems therefore probable that $b^{\rm
electroweak}_{\rm vac} \lesssim 1$, so that the electroweak transition
has only a weak attracting effect on $\varphi$. We conclude that phase
transitions seem to have only a modest effect on $\varphi$. It would be
at present meaningless to refine the calculation of the effect on
$\varphi$ of the electroweak and QCD phase transitions [even the order
of the transitions is in doubt, not to mention the precise redshift
dependence of $\lambda (p)$ and $s_{\rm vac} (p)$]. In fact, until one
has some understanding of the cosmological constant problem, it does
not make much sense to compute any gravitational effect linked to
phase transitions.
 In the following
we shall therefore neglect the effect of the phase transitions with
respect to that of the nine mass thresholds discussed above.

\section{Evolution of the dilaton during the matter-dominated era}
\label{sec:5}

 The matter content of the universe near the end of the radiation era
and during the subsequent matter era can be described as the
superposition of a relativistic gas (``radiation", i.e. photons and
three neutrinos in the standard picture) and of a non-relativistic one
(``matter"; made of particles of mass $m_m (\varphi)$). From
Eqs.~(\ref{eq:3.5}) the source terms for the cosmological evolution
equations (\ref{eq:3.2}) read $\rho =\rho_r +\rho_m$, $P=P_r +P_m$,
$\sigma =-\alpha_m (\varphi) (\rho_m -3P_m)$ with $P_r =\rho_r/3$,
$P_m \simeq 0$, and $\alpha_m (\varphi) \equiv \partial \ln\, m_m
(\varphi) /\partial\varphi$. Either from the definition (\ref{eq:3.5a})
or from the separate energy balance equations discussed below
Eq.~(\ref{eq:3.6}), one deduces that, during the expansion, $\rho_r
\propto a^{-4}$ while $\rho_m \propto m_m (\varphi) a^{-3}$. Finally,
the evolution of $\varphi$ with respect to the $p$-time $p\equiv \ln a
+{\rm const.}$ is given by the equation
\begin{equation}
 {2\over 3-\varphi'^2} \varphi'' + [1-\lambda (p,\varphi)] \varphi' =
 - [1-3\lambda (p,\varphi)] \alpha_m (\varphi)\ , \label{eq:5.1}
\end{equation}
with $3\lambda (p,\varphi) = [1+C m_m (\varphi) e^p]^{-1}$, $C$ being
some constant. In the approximation where the radiation era has already
attracted $\varphi$ very near a minimum $\varphi_m$ of $m_m(\varphi)$, we
can consider that $m_m(\varphi) \simeq {\rm const.}$ in $\lambda
(p,\varphi)$. Choosing now the
origin of $p$ at the equivalence between radiation and matter $[\rho_r
(p=0) =\rho_m (p=0)]$, we get simply $\lambda (p)= 3^{-1} (1+e^p)^{-1}$.
Neglecting $\varphi'^2$ in Eq.~(\ref{eq:5.1}) and using the harmonic
approximation (\ref{eq:4.6}), we find that $\varphi$ satisfies a linear
differential equation which can be rewritten as a hypergeometric
equation. Denoting $x\equiv e^p \equiv a/a_{\rm equivalence}$ we have
\begin{equation}
 x(x+1) \partial^2_x \varphi + \left( {5\over 2} x+2\right) \partial_x
\varphi + {3\over 2} \beta_m (\varphi -\varphi_m) = 0\ . \label{eq:5.2}
\end{equation}
The condition of regularity of $\varphi$ when $x\to 0$, say $\varphi
(x=0) =\varphi_{\rm rad}$ ($\varphi_{\rm rad}$ denoting the value of
$\varphi$ at the end of the radiation era, before the transition to the
matter era around $p=0$), selects uniquely the solution of
(\ref{eq:5.2}) to be $\varphi_m + (\varphi_{\rm rad} -\varphi_m) \times
F[a,b,c;-x]$. Here $F[a,b,c;z]$ denotes the usual (Gauss) hypergeometric
series. The values of the parameters are
\begin{equation}
  a = {3\over 4} -i\omega\ , \quad b={3\over 4} +i\omega\ ,\quad c=2\ ,
 \label{eq:5.3}
\end{equation}
with $\omega \equiv \left[ {3\over 2} \left( \beta_m - {3\over 8}
\right)\right]^{1/2}$. In other words, the attraction factor of the matter
era up to the present time, $F_m\equiv (\varphi_{\rm now} -\varphi_m
)/(\varphi_{\rm rad} -\varphi_m)$, is given by
\begin{equation}
 F_m = F[a,b,c; -Z_0]\ , \label{eq:5.4}
\end{equation}
where $Z_0 \equiv e^{p_0} \equiv a_{\rm now} /a_{\rm equivalence}$
denotes the (Einstein frame) redshift separating us from the moment of
equivalence between matter and radiation. As $Z_0$ is large
(see Eq.(\ref{eq:6.5a}) below), we can use
the asymptotic behavior of the hypergeometric function (together with the
properties of Euler's $\Gamma$ function) to get more explicit forms for
$F_m$. Whatever be the sign of $\beta_m -3/8$ (i.e. in the two cases
where $\omega$ is real or pure imaginary) one can write
\begin{equation}
 F_m = 2^{1/2} \pi^{-1/2} 2^{2i\omega} \Gamma_2 e^{-{3\over 4}p}
   e^{i\omega p} + (i\omega \leftrightarrow -i\omega) \ , \label{eq:5.5}
\end{equation}
with $\Gamma_2 \equiv \Gamma (2i\omega) / \Gamma (2i\omega +3/2)$.
Actually, from the estimate (\ref{eq:4.6b}) we expect $\beta_m$ to be (much)
larger than 3/8 (indeed, if $m_m \sim 1$~GeV, one would need $\kappa$ to be
smaller than $9.2\times 10^{-3}$ to make $\beta_m < 3/8$). In that case
($\omega$ real), one can compute the modulus of the complex number
$\Gamma_2$ in terms of elementary functions to get
\begin{equation}
 F_m = \left[ {{\rm cotanh} (2\pi\omega)\over \pi\omega (\omega^2+1/16)}
 \right]^{1/2} e^{-{3\over 4} p_0} \cos \theta_0 \ , \label{eq:5.6}
\end{equation}
with $\theta_0 \equiv\omega p_0+2\omega \ln 2+ {\rm Arg} (\Gamma_2)$. As in
the case of attraction by mass thresholds (when $\beta_A f^{\rm in}_A
\gtrsim 1$), the attraction factor (\ref{eq:5.6}) is proportional to a
cosine (when $\beta_m > 3/8)$ because Eq.~(\ref{eq:5.1}) describes a
damped oscillation around the minimum $\varphi_m$ of $\ln m_m
(\varphi)$. When $\beta_m < 3/8$, $\varphi$ slowly rolls down toward
$\varphi_m$ without oscillating (overdamped oscillator). [See also
Ref.~\cite{DN} in which the transition between radiation domination
and matter domination was approximated ---~in the analytical
formulas~--- as being a sharp one].

\section{Observable consequences of a cosmological relaxed massless
dilaton.}

 Sections 4 and 5 have exhibited several efficient mechanisms for
driving the VEV of the dilaton toward a value where it decouples from
matter. However, none of these mechanisms is a perfect attractor. The
important question remains of giving quantitative estimates of the
residual coupling strength of the dilaton at various cosmological epochs
and of the corresponding observable effects.

The quantitative estimates of the efficiency of the cosmological
attraction of the dilaton depend very much on the universality, or lack
thereof, of the dilaton couplings. If the dilaton coupling functions
$B_g(\Phi)$, $B_F(\Phi)$, $B_H(\Phi)$ (the latter representing the class
of couplings to the fundamental Higgs sector, if it exists as such) are
unrelated functions, one expects the mass functions (\ref{eq:2.10}) to
have minima (if any) at different values of $\varphi$, say
$\varphi^A_m$. For instance, the lepton and quark masses will involve
$B_H$ while hadron masses will all be proportional to $B_g^{-1/2} \exp
[-8\pi^2 b_3^{-1} k_3 B_3]$ ($B_3 \equiv B_{SU(3)}$). In such a non-universal
case, the various mass thresholds, and phase transitions, will not
attract $\varphi$ to the same value, but will tend to reshuffle each
time the value of $\varphi$. In that case, the only efficient fixing
of the value of $\varphi$ would arise during the matter era, $\varphi$
being attracted toward of minimum of $m_m(\varphi)$ where the label
``$m$'' represents the type of matter which dominates the present
universe.

By contrast, one can  consider the case where all the dilaton
coupling functions coincide, $B_i(\Phi)= B(\Phi)$. This case of
universal coupling of the dilaton to matter has a suggestive simplicity.
It looks like a natural generalization of the universal $e^{-2\Phi}$
coupling arising at the string tree level. In the universal $B(\Phi)$
case, all the mass thresholds, as well as the QCD phase transition and
the matter era, tend to attract $\varphi$ to a common value, some
maximum $\varphi_m$ of $B(\varphi)$.\footnote{Note that a primordial
(inflationary type) phase transition ---~as well as the electroweak one,
if the Higgs sector is fundamental~--- could instead attract $\varphi$
to a \underbar{minimum} of $B(\varphi)$ through a transient vacuum
energy $\propto B(\varphi)$.} In the universal case, the cosmological
evolution is an extremely efficient way of pinning down the value of
$\varphi$. Moreover, as $\varphi$ is pinned down to an extremum of
$B(\varphi)$, i.e. to a value where $\partial B(\varphi) /\partial
\varphi$ and $\partial m_A (\varphi) /\partial \varphi$ vanish, one can
say that the universal dilaton coupling case illustrates some
``Principle of Least Coupling'' in the sense that the universe is
attracted to dilaton values extremizing the strengths of the interaction.
It would be worth exploring whether
imposing this universality provides a sensible way of selecting a
preferred class of string models. In the following, we leave open the
two possibilities, universal/non-universal, in our discussion of the
observable consequences of our scenario.

The earliest observational information we have about cosmology concerns
the primordial abundance of the light elements (mainly Helium 4, with
traces of Deuterium, Helium 3 and Lithium 7). Let us discuss the
production of Helium 4 as an example. In the standard scenario of
homogeneous primordial nucleosynthesis, the abundance of Helium is
mainly determined by the neutron/proton ratio at the temperature where
the rate of interconversion $n\leftrightarrow p$ due to weak interactions
becomes slower than the cosmological expansion rate (freeze-out)
(see Ref.~\cite{KT}).
Neglecting the small additional effect of free neutron decay, one can write
an approximate analytical formula for the primordial Helium abundance (by
weight), $Y$, of the form, $Y=2/(\exp (aX)+1)$ where $a$ is a pure
number of order unity and where $X$ denotes the following dimensionless
combination of coupling constants and masses
\begin{equation}
  X \equiv g^{4/3}_2 (1+3g^2_A)^{1/3} \left( {m_n-m_p\over m_W}\right)
 \left({m_{\rm Planck}\over m_W}\right)^{1/3} \left({g_*\over 10.75}\right)
 ^{-1/6}\ . \label{eq:6.1}
\end{equation}
Here $g_2$ denote the $SU(2)$ coupling constant, $g_A\simeq 1.26$ the
axial/vector coupling of the nucleon, and $g_*$ the effective number of
relativistic degrees of freedom at freeze-out (retained here to allow
easy comparisons between the effect of a change in $g_*$ ---~e.g. an
additional light neutrino~--- and the effects of changing, e.g.,
Newton's constant $G=m^{-2}_{\rm Planck}$, or Fermi's one $G_F = g^2_2
/8m^2_W$). A remarkable fact about the combination $X$ is that it is
numerically of order unity thanks to a delicate compensation between
large $\left[ (m_{\rm Planck} /m_W)^{1/3} \simeq (1.52 \times 10^{17})
^{1/3}\right]$ and, small $[(m_n-m_p)/m_W \simeq 1.61\times 10^{-5}]$
factors. This fact prevents us from proposing an educated guess of the
quantitative dependence of $X$ on the bare dilaton coupling constants
$B_F(\varphi)$, $B_g(\varphi)$,\dots . Even the sign of $\partial\ln X/
\partial\ln B^{-1}$ (when $B_i(\varphi) = B(\varphi)$) is unclear.
On the other hand one can estimate that $\partial Y/\partial \ln X \simeq
-0.44$ both from the rough analytical formula for $Y(X)$ and from the
numerical computations of the dependence of $Y$ on the neutron half-life
or on $g_*$. We can therefore write the value of the Helium abundance
predicted by a scenario modified by the presence of a dilaton as
\begin{equation}
 Y^{\rm dil}(\eta) = Y^{\rm GR} (\eta) - 0.22 {\partial\ln X\over
\partial\ln B^{-1}}\kappa (\varphi_{\rm rad}-\varphi_m)^2\ ,\label{eq:6.2}
\end{equation}
where we have reestablished the slight dependence of $Y$ upon the baryon
to photon ratio, $\eta$. In the standard, general relativistic scenario
the dependence of the GR-predicted abundances on $\eta$ is crucially
used, together with the observed values of the light-element abundances,
to set upper bounds on
$\eta$, and thereby upper bounds of the ratio of the present total
baryon mass density to the closure density, $\Omega_b$. The standard
conclusion being that baryons fail to close the universe by at least
a factor five, $\Omega_b < 0.2$\cite{KT}. Eq.~(\ref{eq:6.2})
 [to be completed by
the corresponding dilaton-modified predictions for the other light
elements] suggest that a dilatonic universe could naturally accomodate
$\Omega_b =1$ if the value $\varphi_{\rm rad}$ of $\varphi$ at freeze-out
(i.e. just after the electron mass threshold) differs by a small (but
not too small) amount from the minimum $\varphi_m$ [For instance, in the
case of the Helium abundance, the dilaton correction term on the
right-hand side of Eq.~(\ref{eq:6.2}) should be approximately $-0.03$,
and $\partial \ln X/\partial\ln B^{-1}$ should be positive]. It would
be interesting to reexamine in full numerical detail primordial
nucleosynthesis within the type of dilaton scenario considered here to
assess whether it could naturally reconcile $\Omega_b =1$ with the
observed abundances of light elements. Let us only note here that the
rather modest attraction toward $\varphi_m$ which is probably needed
in such a scenario seems more natural in the non-universal case. Indeed,
in the universal $B(\varphi)$ case, all the nine mass thresholds
compound their effect to drive $\varphi$ very near some universal
minimum $\varphi_m$.  More precisely, $\varphi_{\rm rad} -\varphi_m =
F_r\times (\varphi_{\rm in} -\varphi_m)$ where $\varphi_{\rm in}$ is the
``initial" value of $\varphi$ (meaning in this work, before the
electroweak phase transition)and where the total attracting power of the
radiation era is given by
\begin{equation}
 F_r (\kappa) =\left\{ \prod^9_{A=1} m_\pm (\beta_A f^{\rm in}_A) \right\}
 \times \left\{ \prod_{i=2,3} p (\beta^{\rm vac}_i f^{\rm vac}_i) \right\}
 \ . \label{eq:6.3}
\end{equation}
 The values of $\beta_A$ and $f^{\rm in}_A$ to be used in the attraction
factors of each of the nine mass thresholds have been given above
[remember that the $+ (-)$ sign corresponds to fermions (bosons)]. The
second factor in Eq.~(\ref{eq:6.3}) corresponds to the effect of the
two known phase transitions electroweak (2) and QCD (3). In view of
the uncertainty in the calculation of the effect of phase transitions,
and anyway of their expected modest contribution (see above), we shall
neglect the attraction power of these phase transitions in the
following. The small but non-zero value of $\varphi_{\rm rad}
-\varphi_m = F_r (\kappa) \Delta \varphi$ [with $\Delta \varphi \equiv
\varphi_{\rm in} -\varphi_m$] implies that all the gauge coupling
constants squared, $g^2 \propto B^{-1} (\varphi)$, differed, at the end
of radiation era, from their present values $g^2_0$ by a fractional
amount
\begin{equation}
 {g^2_{\rm rad} -g^2_0\over g^2_0} \simeq {1\over 2} \kappa
 (\varphi_{\rm rad}-\varphi_m)^2 = {1\over 2} \kappa (F_r (\kappa)
  \Delta\varphi)^2  \label{eq:6.4}
\end{equation}
[where we used the fact that $\varphi_0 -\varphi_m \ll \varphi_{\rm rad}
-\varphi_m$ because of the matter era attraction]. As one \underbar{a
priori} expects $\Delta\varphi \equiv \varphi_{\rm in} -\varphi_m$ to
be of order unity, the function ${1\over 2} \kappa F_r^2 (\kappa)$, which
is plotted in Fig.~2, illustrates the remarkable efficiency (in the
universal case) of the radiation era in pinning down the values of the
physical coupling constants.

During the subsequent matter era, $\varphi$ is (in the universal
case) further driven toward
$\varphi_m$ by the factor $F_m(\kappa, Z_0)$,
Eqs.~(\ref{eq:5.4})-(\ref{eq:5.6}). The numerical value of the matter-era
attraction factor $F_m$ is proportional to $Z^{-3/4}_0$  where
$Z_0\equiv e^{p_0}$ denotes the redshift separating us from the epoch of
equivalence between matter and radiation. In the approximation $m_m
(\varphi) \simeq {\rm const.}$ introduced at the beginning of
Sec.~\ref{sec:5}, this redshift is given by \cite{DN}
\begin{mathletters}
\label{eq:6.5}
\begin{equation}
 Z_0 = {\rho^{\rm matter}_0 \over \rho^{\rm rad}_0} \simeq 13350\
 \Omega_{75}\ , \label{eq:6.5a}
\end{equation}
where
\begin{equation}
 \Omega_{75} \equiv 8\pi \overline{G}\, \rho^{\rm matter}_0 /
[3 (75\ \hbox{km s}^{-1}\hbox{Mpc}^{-1})^2] = \rho^{\rm matter}_0 /
 1.0568 \times 10^{-29} \hbox{g cm}^{-3} \ . \label{eq:6.5b}
\end{equation}
\end{mathletters}
Under the hypothesis of a spatially flat universe $(K=0)$, generally
assumed in this paper, $\Omega_{75}$ is linked to the present value of
Hubble's ``constant", $H_0$, by
$\Omega_{75}= (H_0 /75$~km~s$^{-1}$Mpc$^{-1})^2$. In that case, the
observational limits $50< H_0/1$~km~s$^{-1}$Mpc$^{-1} < 100$ imply
$0.44<\Omega_{75} <1.78$. On the other hand, if one assumes that the
universe is spatially hyperbolic $(K=-1)$, one must modify the
coefficients of the evolution equation (\ref{eq:3.3}) for $\varphi$ by
retaining the $K$-dependent terms. However, it was shown in
\cite{DN} that as long as $\Omega_{75} >0.05$ this modification of
Eq.~(\ref{eq:3.3}) has a small effect, and that the matter-era
attraction factor of $K=-1$ universes is well approximated by the $K=0$
formula (\ref{eq:5.6}), with $Z_0$ given by Eqs.~(\ref{eq:6.5a}),
(\ref{eq:6.5b}). The main difference is that now $\Omega_{75}$ is not
related to $H_0$, and can be smaller than 0.44. In fact, present
observational data are compatible with $\Omega_{75} \sim 0.1$.

Finally, the scenarios considered here predict that the present value
of $\varphi$, say $\varphi_0$, differs from the minimum $\varphi_m$
by $\varphi_0 - \varphi_m = F_t (\kappa, Z_0) \Delta \varphi$ where
$\Delta\varphi \equiv \varphi_{\rm in} -\varphi_m$ and where the total
attraction factor is
\begin{equation}
 F_t (\kappa, Z_0)\equiv F_r (\kappa)\,F_m (\kappa, Z_0)\ .\label{eq:6.6}
\end{equation}
There are three kinds of presently observable consequences of having
$\varphi_0$ near, but different from, $\varphi_m$: (i) violations of the
(weak) equivalence principle; (ii) modifications of relativistic gravity,
and; (iii) slow changes of the coupling constants of physics, notably
the fine-structure constant $\alpha$ and Newton's constant $G$.

To discuss the modifications of the gravitational sector, we can make
use of the results of Ref.~\cite{DEF1} on the relativistic
gravitational interaction of condensed bodies in generic
metrically-coupled tensor-scalar theories. Indeed, the action describing
the classical interaction of massive particles of various species under
the exchange of the $g_{\mu\nu}$ and $\varphi$ fields is given by
$S_{g,\varphi} + S_m [g,\varphi,x]$ where $S_{g,\varphi}$ is given by
(\ref{eq:2.7b}) and $S_m$ by  Eq.~(\ref{eq:3.4}). This action is
identical to the one studied in Sec.~6 of \cite{DEF1}. We conclude
that, at the Newtonian approximation, the interaction potential between
particle $A$ and particle $B$ is $-G_{AB} m_A m_B /r_{AB}$ where
$r_{AB} \equiv |{\bf x}_A -{\bf x}_B|$ and
\begin{equation}
  G_{AB} = \overline G (1 + \alpha^{(0)}_A \alpha^{(0)}_B)\ .
  \label{eq:6.7}
\end{equation}
Here $\overline G$ is the bare gravitational coupling constant entering
the action (\ref{eq:2.7b}), and $\alpha^{(0)}_A$ is the present strength
of the coupling of the dilaton to $A$-type particles, i.e. the value of
(\ref{eq:3.6}) taken at the cosmologically determined VEV $\varphi_0$.
[In diagrammatic language, the two terms on the right-hand side of
Eq.~(\ref{eq:6.7}) are, respectively, the one-graviton exchange
contribution $(\overline G)$ and the one-dilaton exchange one $(\overline
G \alpha^{(0)}_A \alpha^{(0)}_B)$]. Two test masses, made respectively
of $A$- and $B$-type particles, will fall in the gravitational field
generated by an external mass $m_E$ with accelerations $a_A$ and $a_B$
differing by
\begin{equation}
 \left( {\Delta a\over a}\right)_{AB} \equiv 2\, {a_A-a_B\over a_A+a_B}
 = {(\alpha^{(0)}_A- \alpha^{(0)}_B) \alpha^{(0)}_E\over 1 + {1\over 2}
 (\alpha^{(0)}_A+ \alpha_B^{(0)}) \alpha^{(0)}_E} \simeq
 (\alpha^{(0)}_A -\alpha^{(0)}_B) \alpha^{(0)}_E \ . \label{eq:6.8}
\end{equation}
All precision tests of the gravitational interaction  used
macroscopic bodies made of (neutral) atoms. Let the labels $A$, $B$,\dots
denote some atoms. In the approximation where one neglects $m_u/m_N$,
$m_d/m_N$, $m_e/m_N$, $\alpha$ and $\alpha_{\rm weak}$, the mass of an atom
is a pure (dilaton-independent) number times a \underbar{QCD-determined}
mass scale, say $u_3(\varphi)$.
In this approximation $\alpha_A(\varphi) =\partial
\ln m_A/\partial\varphi$ is \underbar{independent} of the type of atom
considered and is equal to $\alpha_3(\varphi)=\partial\ln u_3/\partial
\varphi$. The dilaton dependence of $u_3$ is determined by
Eq.~(\ref{eq:2.9}). Choosing $u_3$ so that its present value
$u_3(\varphi_0)$ is numerically equal to the atomic mass unit,
$u=931.49432$~MeV, we see from Eqs.~(\ref{eq:4.6}) that
\begin{equation}
 \alpha_A (\varphi_0) \simeq \alpha_3 (\varphi_0) = \beta_3 (\varphi_0
 -\varphi_m) =\beta_3 F_t (\kappa, Z_0) \Delta \varphi\ , \label{eq:6.9}
\end{equation}
with $\beta_3 =\kappa \partial\ln u_3 /\partial\ln B^{-1} \simeq 40.82\,
\kappa$.

 In this approximation, the dilaton mimics a usual Jordan-Fierz
(-Brans-Dicke) field, i.e. a scalar field coupled exactly to
$T^\mu_\mu$. The main observational consequences of the body-independent
coupling (\ref{eq:6.9}) are modifications of \underbar{post-Newtonian}
relativistic effects, $O(v^2/c^2)$ beyond the Newtonian $1/R$ interaction
(weak gravitational field case )\footnote{In view of the positiveness of
$\beta_3$, the recent results of \cite{DEF2} show that the deviations
from general relativity are further quenched in the
strong-gravitational-field case of binary neutron star systems.}.
The latter are measured by the two
Eddington parameters $\gamma_{\rm Edd}-1$ and $\beta_{\rm Edd}-1$ (which
vanish in general relativity). From \cite{DEF1} we see that in the
approximation (\ref{eq:6.9})
\begin{equation}
 1 - \gamma_{\rm Edd} = 2 {\alpha^2_3\over 1+\alpha^2_3} \simeq 2
(\beta_3)^2 (F_t (\kappa, Z_0) \Delta\varphi)^2\ , \label{eq:6.10}
\end{equation}
\begin{equation}
 \beta_{\rm Edd}-1 = {1\over 2} {\beta_3 \alpha^2_3\over
\left(1+\alpha^2_3\right)^2} \simeq {1\over 2}
(\beta_3)^3 (F_t (\kappa, Z_0) \Delta\varphi)^2\ . \label{eq:6.11}
\end{equation}
Note also that the value of Newton's gravitational constant
(in Einstein units) is $G_N = \overline G (1+\alpha^2_3)$.

 Much more sensitive tests of the existence of dilaton couplings are
obtained by looking at violations of the weak equivalence principle,
i.e. at the body-dependence of $\alpha_A(\varphi_0)$ beyond the QCD
approximation (\ref{eq:6.9}). To do this, we shall retain the leading
$m_u/m_N$, $m_d/m_N$, $m_e/m_N$ and $\alpha$ corrections to the
mass of an atom. First, the mass of the nucleons have the form, $m_p
=m_{N3} +b_um_u +b_dm_d+C_p\alpha$, $m_n=m_{N3}+b_dm_u+b_um_d
+C_n\alpha$, where $m_{N3}(\simeq u_3)$ is the pure QCD approximation
to the nucleon mass, and where $b_u$, $b_d$, $C_p/u_3$ and $C_n/u_3$ are
pure numbers (in the approximation of negligible strange-quark content
one has $b_u=\langle p|\overline uu|p\rangle /2m_N$, $b_d= \langle p|
\overline dd|p\rangle/2m_N$)\cite{GL}. Second, the mass of an atom can be
approximately decomposed as
\[ m({\rm Atom}) = Z m_p + Nm_n + Zm_e + E^{\rm nucleus}_3
   + E^{\rm nucleus}_1 \ , \]
where $Z$ is the atomic number and $N$ the number of neutrons, and where
$E^{\rm nucleus}_3$ denotes the strong-interaction contribution to
the binding energy of the nucleus, and $E^{\rm nucleus}_1$ the Coulomb
interaction energy of the nucleus.

In terms of the baryon number $B\equiv N+Z$, the neutron excess
$D\equiv N-Z$, and the Coulomb energy term $E\equiv Z(Z-1) /(N+Z)^{1/3}$,
the mass of an atom can be written as,
\begin{equation}
  m({\rm Atom}) = u_3 M_3 +\sigma' B+ \delta' D + a_3 \alpha u_3 E\ ,
 \label{eq:6.12}
\end{equation}
where $M_3$ is a pure number ($=B$+ strong-interaction binding contribution)
and where  we have defined
\[ \sigma' \equiv \sigma +{1\over 2} C_n \alpha + {1\over 2} C_p \alpha
 + {1\over 2} m_e\ ,\quad \delta' \equiv {1\over 2}\delta +{1\over 2}
 C_n\alpha - {1\over 2} C_p \alpha - {1\over 2} m_e \ , \]
with the usual definitions for $\sigma \equiv {1\over 2} (m_u+m_d)(b_u+b_d)$,
$\delta \equiv (m_d -m_u)(b_u-b_d)$. [Note the factor $1/2$ in the first
term of the definition of $\delta'$]. Finally, by differentiating the
logarithm of (\ref{eq:6.12}) we get a more precise expression than
(\ref{eq:6.9}) for the dilaton coupling strength
\begin{equation}
 \alpha_A (\varphi_0) \simeq \alpha_3 (\varphi_0) +
{\partial\widehat\sigma \over \partial\varphi_0} \left({B\over M}\right)_A
+ {\partial\widehat\delta \over \delta\varphi_0} \left({D\over M}\right)_A
+ a_3 {\partial\alpha\over \partial\varphi_0} \left({E\over M}\right)_A\ ,
\label{eq:6.13}
\end{equation}
where we have introduced  $\widehat\sigma \equiv \sigma'/u_3$,
$\widehat\delta \equiv \delta'/u_3$ and approximated $M_3\simeq M\equiv
m{\rm (Atom)}/u_3$ in the corrections terms.\footnote{Note that the
assumption of a universal $B(\varphi)$ is crucial to ensure that all the
terms in Eq.(\ref{eq:6.13}) have in common a very small factor
$\varphi_0 - \varphi_m$. If, e.g., the mass of leptons (and/or $\alpha$)
depended on a different function of $\varphi$ than the mass of hadrons,
Eq.(\ref{eq:6.13}) would, at best, predict that the equivalence principle
is violated at the (unacceptable) level
$\sim ( O(\alpha) + O(m_{\rm lepton}/m_{\rm N}) )^2,$ in the favourable
case where the universe is assumed to be dominated by hadronic matter.}
Finally, from Eq.~(\ref{eq:6.8}) we get an equivalence-principle violation
of the form
\begin{equation}
 \left( {\Delta a\over a}\right)_{AB} = (\kappa F_t (\kappa, Z_0) \Delta
 \varphi )^2 \left[ C_B \Delta \left( {B\over M}\right) + C_D \Delta
 \left( {D\over M} \right) + C_E \Delta \left( {E\over M}\right) \right]
 _{AB} \ , \label{eq:6.14}
\end{equation}
where $(\Delta X)_{AB} \equiv X_A -X_B$ and where
$C_B =\lambda_{u_3} \partial \widehat\sigma /\partial\ln B^{-1}$,
$C_D =\lambda_{u_3} \partial \widehat\delta /\partial\ln B^{-1}$,
$C_E = \lambda_{u_3} \lambda_\alpha a_3 \alpha$, $\lambda_{u_3} \equiv
\partial\ln u_3/\partial \ln B^{-1}$ and $\lambda_\alpha \equiv \partial
\ln \alpha /\partial\ln B^{-1}$. Numerically, our usual estimate
(\ref{eq:2.9}) gives $\lambda_{u_3} \simeq 40.82$, and the idea of
unification of gauge couplings at the string scale gives $\lambda_\alpha
\simeq 1$. [E.g. in the simplest $SU(5)$-type GUT the value of the fine
structure constant at the QCD-confining energy scale $u_3$ ---~such that
$\alpha_{\rm strong} (u_3) \simeq 1$~--- is given by
\[ \alpha (u_3) ^{-1} = (22/7) \alpha_{\rm GUT}^{-1}
   - (10/21) \alpha_{\rm strong} (u_3)^{-1}
   \simeq (22/7) \alpha_{\rm GUT}^{-1} \propto B(\varphi) ]\ . \]
We have also $a_3\alpha = 0.717$ MeV$/u_3 =0.770 \times 10^{-3}$ from
the fit of atomic masses to the Bethe-Weizs\"acker formula. We can
therefore estimate the coefficient of the nuclear Coulomb energy term in
Eq.~(\ref{eq:6.14}) to be $C_E \simeq 3.14\times 10^{-2}$. As for the
other two coefficients, $C_B$ and $C_D$, it is much less clear how to
estimate them. From the experiment-derived values of $\sigma =35\pm
5$~MeV and $\delta = 2.05 \pm 0.30$~MeV, plus the theoretical estimates
$C_p \alpha = 0.63$~MeV, $C_n\alpha =-0.13$~MeV \cite{GL}, one can compute
$\widehat \sigma =3.8\times 10^{-2}$ and $\widehat\delta = 4.2\times
10^{-4}$. From the point of view of their dilaton dependence
$\widehat\sigma$ and $\widehat\delta$ are the sum of four terms
proportional to $m_u/u_3$, $m_d/u_3$, $m_e/u_3$ and $\alpha$. It is
impossible at present to reliably guess the dilaton-dependence of the
mass ratios $m_{\rm quark}/m_{\rm hadron}$ and $m_e/m_{\rm hadron}$.
The numbers we would get for $\partial\widehat\sigma/\partial \ln
B^{-1}$ would be very different were we to assume our usual exponential
link to the string scale, or some other assumption. It seems however
reasonable to estimate that the order of magnitude of $\partial\widehat
\sigma /\partial\ln B^{-1}$ and $\partial\widehat\delta /\partial\ln
B^{-1}$ will be \underbar{at most} that given by the exponential
assumption (\ref{eq:2.11}), and \underbar{at least} that obtained by
differentiating only the fine-structure constant contributions to
$\widehat\sigma$ and $\widehat \delta$. This yields corresponding rough
upper and lower bounds for the coefficients of the $B$ and $D$
contributions: $1.1\times 10^{-2} \lesssim |C_B|\lesssim 7.5$,
$1.7 \times 10^{-2}  \lesssim |C_D|\lesssim 8.2\times 10^{-2}$.
If these upper bounds are
correct, one can check that the last term in Eq.~(\ref{eq:6.14}) will be
numerically dominant for pairs $(A,B)$ having a large difference in
atomic number [Indeed, $E/M$ is roughly proportional to $Z^{2/3}$].
The largest effect would arise in comparing Uranium
(for which $E/M \simeq 5.7$) with Hydrogen (any
light element would do nearly as well). For such a pair,
Eq.~(\ref{eq:6.14}) yields
\begin{equation}
 (\Delta a/a)_{\rm max} = 0.18 (\kappa F_t (\kappa, Z_0)
\Delta\varphi)^2 \ . \label{eq:6.15}
\end{equation}
The right-hand side of Eq.~(\ref{eq:6.15}) is plotted in Fig.~3 as a
function of $\kappa$ (assuming $\Delta\varphi =1$, and $\Omega_{75}=1$).
As one \underbar{a priori} expects $\kappa$ to be of order unity, Fig.~3
shows that, within the scenario considered here
(including universal dilaton couplings), the present tests of
the equivalence principle (at the $10^{-11}-10^{-12}$ level) do not put
any significant constraints on the existence of a massless dilaton.

The situation is even worse if we consider tests of post-Newtonian gravity.
Indeed, from Eqs.~(\ref{eq:6.10}) and (\ref{eq:6.15}) we have the link
\begin{equation}
 \left( {\Delta a\over a}\right)_{\max} \simeq 5.4 \times 10^{-5}
 (1 - \gamma_{\rm Edd}) \label{eq:6.16}
\end{equation}
showing that the present and planned levels of testing of post-Newtonian
gravity, i.e. $10^{-3}$ and $10^{-7}$ at best for $\gamma_{\rm Edd}$,
correspond, respectively, to equivalence-principle tests at the levels
$5\times 10^{-8}$ and $5\times 10^{-12}$. The other link $\beta_{\rm Edd}
-1 ={1\over 4} \beta_3 (1-\gamma_{\rm Edd}) \simeq 10.2 \kappa
(1-\gamma_{\rm Edd})$ shows that $\beta_{\rm Edd}$ tests do not fare
essentially better.

The last observational consequence of our scenario to discuss is the
residual present variation of the coupling constants of physics. From
$\partial\ln\alpha /\partial\ln B^{-1} \simeq 1$, with $\ln B^{-1} =
{\rm const.} + {1\over 2} \kappa (\varphi (t) -\varphi_m)^2$ and a
present time dependence of $\varphi$ given by Eq.~(\ref{eq:5.6}) [in
which the leading term is $\cos (\omega p+{\rm const.})$], we deduce
that
\begin{equation}
 \left( {\dot\alpha\over \alpha H}\right)_0 = - \kappa
\left[ \omega \tan \theta_0 + {3\over 4} \right]
 (F_t (\kappa, Z_0) \Delta\varphi)^2\ .\label{eq:6.17}
\end{equation}
Using $\omega \equiv [{3\over 2} (\beta_m - {3\over 8})]^{1/2}$ with
$\beta_m\sim 40.8\kappa$ (if the particles dominating the universe have a
mass not very different from the GeV scale), we have the approximate link
$(\dot\alpha/\alpha H)_0 \sim -43 \kappa^{-1/2} \tan \theta_0 (\Delta
a/a)_{\max}$. This link shows again that equivalence principle tests are
the most sensitive way of searching for possible dilaton couplings
[In the foreseeable future, ultrastable cold-atom clocks might probe the
level $\dot\alpha/\alpha \sim 10^{-16}$yr$^{-1} \sim 10^{-6}H_0$ which
corresponds to $\Delta a/a \sim 10^{-8}$].

 Finally, the time variation of the gravitational coupling
constant (measured in Einstein units) is
\begin{equation}
 \left( {\dot G\over GH}\right)_0 = - 2
\left[ \omega \tan \theta_0 + {3\over 4} \right]
(\beta_3 F_t (\kappa, Z_0) \Delta \varphi)^2\ . \label{eq:6.18}
\end{equation}
In actual $\dot G$ experiments one is comparing an orbital frequency
$n$ (e.g. let us consider that of a planet around the Sun) to an
atomic frequency $\nu$. Taking into account the adiabatic invariants
of the orbital motion (angular momentum and eccentricity) and assuming
an atomic clock based on the Bohr frequency $\propto m_e \alpha^2$, the
directly measured quantity will be
\begin{equation}
 {\dot n\over n} -{\dot \nu\over \nu} = 2 {\dot G\over G} + 2{\dot
m_s\over m_s} + 3 {\dot m_p\over m_p}
  - {\dot m_e\over m_e} - 2 {\dot\alpha\over \alpha} \ ,
  \label{eq:6.19}
\end{equation}
where $m_s$, $m_p$ denote the  masses of the Sun and of the planet.
[Contrary to Eq.~(\ref{eq:6.18}), Eq.~(\ref{eq:6.19}) is valid in any
system of units].
 Eq.~(\ref{eq:6.19}) gives
finally\footnote{The link between $\dot \varphi_0$ and ${\dot n} / n$
is more involved if $n$ is the orbital frequency of a binary neutron
star system: see \cite{N}, which must be completed by taking into
account the changes in the rest-masses of the stars, and the
non-perturbative gravitational self energy effects \cite{DEF2}.}
\begin{equation}
 {\dot n\over n} -{\dot\nu\over \nu} =-  H_0
\left[ \omega  \tan \theta_0 + {3\over 4} \right] (F_t
(\kappa, Z_0) \Delta\varphi)^2 \left[ 4\beta^2_3 + 5 \beta_3
  - \beta_e - 2 \kappa \right]\ , \label{eq:6.20}
\end{equation}
in which the term coming from Eq.~(\ref{eq:6.18}) dominates. In spite
of the large factor $(\beta_3/\kappa)^2\simeq (40.8)^2$, the scaling
of the prediction (\ref{eq:6.20}) with the Hubble rate $H_0$ makes it
pale in comparison with equivalence principle tests. [On the other hand,
this large factor renders $\dot G$ experiments competitive with
$\dot\alpha$ ones].

\section{Conclusions}

 Einstein's starting point in constructing general relativity was the
interpretation of the universality of free fall in terms of a universal
coupling of matter to a common metric tensor $g_{\mu\nu}$. It has since
been felt that such a universal metric coupling was the only
theoretically natural way of explaining how the long-range fields
participating in gravity\footnote{In Einstein's theory gravity is
mediated by only one, spin 2, field; but in metrically-coupled
tensor-scalar theories, gravity is mediated both by a spin 2 and a spin
0 field. In the latter case, the universal metric coupled to matter is a
combination of the two pure-spin fields: $g^{\rm univ}_{\mu\nu} = A^2
(\varphi) g^*_{\mu\nu}$.} could satisfy the high-precision tests of the
equivalence principle (now reaching the $10^{-12}$ level).

The present work suggests that a universal multiplicative coupling of a
long-range scalar field $\Phi$ to all the other fields, ${\cal L}_{\rm
tot} = B(\Phi) {\cal L}_0 (g_{\mu\nu}, \Phi, A_\mu, \psi,\dots)$
with $B(\Phi)$ admitting a local maximum, though a priori entailing
strong violations of the equivalence principle, provides another
theoretically natural way of explaining why no violations have been seen
at the $10^{-12}$ level. It maybe worthwhile to summarize in qualitative
terms\footnote{Simplified quantitative estimates are provided below.}
the basic reasons why a massless dilaton is rendered nearly
invisible during the cosmological evolution: (i) Each time, during
the radiation era, the universe passes through a temperature $T \sim
m_A$ the $A$-type particles and antiparticles become nonrelativistic
before annihilating themselves and disappearing from the cosmic soup;
this provides a source term for the dilaton proportional to the
 $\varphi$-gradient of $m_A(\varphi)$, which attracts $\varphi$ toward
a minimum $\varphi_m^A$ of $m_A(\varphi)$; Eq.~(\ref{eq:4.6b}) and
Fig.1 suggest that each such attraction is moderately efficient, leaving
$\varphi$ nearer to $\varphi_m^A$ by a factor $\sim {1/3}$.
(ii) Under the assumption of universality of the dilaton coupling
functions $B(\varphi)$, the minima of all the mass functions
$m_A(\varphi)$ will coincide and the $\sim$ 9 mass thresholds of
the radiation era will compound their effects to attract
very efficiently  $\varphi$ toward
some common minimum $\varphi_m$.
(iii) In the subsequent matter era, $\varphi$ will be continuously
attracted toward a minimum of the mass function $m_m(\varphi)$
corresponding to the (nonrelativistic) matter dominating the universe.
[ Under the same universality condition this minimum will be again
$\varphi_m$.] The attraction factor due to the matter era is inversely
proportional to the 3/4th power of the redshift $Z_0 \sim 1.3 \times 10^4$
separating us from the end of the radiation-dominated era.
 (iv) As a consequence of the very efficient total attraction toward
$\varphi_m$, the present strength of the coupling of the dilaton to
any type of matter $\alpha_A(\varphi)$ , being proportional to the
$\varphi$-gradient of $m_A(\varphi)$, is very small. The present
deviations from general relativity in the interaction between two masses,
$m_A$ and $m_B$, are proportional to the product $\alpha_A \alpha_B$
and are therefore extremely small.
(v) The equivalence principle tests are very sensitive, but they probe
only differences $(\alpha_A - \alpha_B) \alpha_C$ which, because of the
known universal features of QCD-generated masses, contain as supplementary
small parameters either the ratio of the quark masses to the nucleon mass,
or the fine-structure constant.

 From a theoretical point of view, our work suggests a criterion for
selecting a preferred class of string models: namely those where
string-loop effects preserve the universal multiplicative coupling
present at tree-level, with a dilaton-dependent function admitting a
local maximum\footnote{The existence of a local maximum in $B(\Phi)$ is
necessary. For instance, in the case of the tree-level coupling function
$\exp (-2\Phi)$, $\Phi$ would continuously roll toward $-\infty$ during
the cosmological expansion, and worse, by Eq.~(\ref{eq:4.4}), $\Phi$
would cause deviations from general relativity, including violations
of the equivalence principle, of order unity or more.}.
It will take, however, an improvement in our
current understanding of supersymmetry breaking in string theory
to see whether the universality required by the Least Coupling
Principle is a viable option, providing a reasonable selection
criterion for SUSY breaking mechanisms.
It is to be noted that in this paper we had always in mind the coupling
of the (four dimensional) dilaton which is such an intimate partner of
the graviton that it seems reasonable to assume that it remains massless
in the low-energy world\footnote{It is enticing to assume that the
presently obscure mechanism ensuring the vanishing of the cosmological
constant allows both `gravitational' fields to remain long-ranged in the
low-energy world.}.
 However, the cosmological attractor mechanism
described here could also apply to the other gauge-neutral scalar fields
(moduli) present in string theory.  Because of threshold effects, the
gauge coupling function $B_F$ acquires a non trivial dependence on the
moduli fields \cite{DKL}. Therefore, we have here a possible mechanism
for fixing the moduli to values where they decouple from the other
fields.

 From an experimental point of view, our results provides a new
incentive to improving the precision of equivalence principle tests
(universality of free fall, constancy of the constants,\dots). Fig.~3
suggests, when assuming that the curvature of $B(\varphi)$ near its
maximum is of order unity ---~say $0.1 < \kappa < 10$~---, to look for a
present level of violation of the universality of free fall somewhere
between $10^{-14}$ and $10^{-23}$. Actually, one should not consider the
results plotted in Fig.~3 too seriously. On the one hand, even within
the precise assumptions made in the text, the predicted maximal value of
$\Delta a/a$ contains an unknown factor $\simeq \Omega^{-3/2}_{75}
(\varphi_{in}-\varphi_m)^2$ which could be $\gtrsim 10$. On the other
hand, our assumption (\ref{eq:2.11}) with $\mu_A =1$ has entailed a
specific phasing of the various oscillations undergone by $\varphi$
during the radiation era, i.e. specific choices of where on the curves
of Fig.~1 $(\varphi (p) -\varphi_m)/(\varphi (-\infty)-\varphi_m)$ ends
up being when $p\to +\infty$. It is possible that the assumption
(\ref{eq:2.11}) has overestimated the combined attraction power of the
radiation era mass thresholds. A different estimate is obtained by
multiplying the WKB approximations (\ref{eq:4.10a}) (all valid as soon
as $\kappa \gtrsim 1$), assuming that all the oscillation angles
$\theta^A_\pm$ are randomly distributed on the circle. Under the latter
assumption it makes sense to compute a $rms$ value of the radiation era
attraction factor $(\langle \cos \theta^A_\pm \rangle_{rms} =1/\sqrt 2)$.
Neglecting as above the effect of the phase transitions one finds
$[F_r (\kappa)]_{rms} = 1.87 \times 10^{-4} \times \kappa^{-9/4}$ for
the radiation era attraction factor, and from Eq.~(\ref{eq:5.6}) with
$\omega \gg 1$, $\langle \cos\theta_0\rangle_{rms} = 1/\sqrt 2$ and
$\Omega_{75} =1$,
$[F_m (\kappa)]_{rms} = 1.47 \times 10^{-5} \times \kappa^{-3/4}$ for
the matter era attraction factor. This leads to a total attraction factor
$[F_t (\kappa)]_{rms} = 2.75 \times 10^{-9} \times \kappa^{-3}$ and
to the following analytical estimate of the $rms$ value of the maximum
value of the equivalence principle violation
\begin{equation}
 (\Delta a/a)^{\max}_{rms} = 1.36 \times 10^{-18} \kappa^{-4}
  (\Delta\varphi)^2\ . \label{eq:7.1}
\end{equation}
The comparison of Fig.~3 with Eq.~(\ref{eq:7.1}) (valid if $\kappa
\gtrsim 1$ and the angles $\theta^A_\pm$ are randomly distributed) indicates
that the phasing of the radiation era oscillations tends to be
destructive. It is possible that alternative assumptions, different from
(\ref{eq:2.11}), yields values of $(\Delta a/a)$ nearer to the $rms$
analytical estimate (\ref{eq:7.1}). This would have the consequence
that presently planned satellite tests of the equivalence principle
\cite{Blaser} which aim at the level $\Delta a/a \sim 10^{-17}$, would
probe a larger domain of values of $\kappa$, $\Delta \varphi$ and
$\Omega_{75}$.

 In conclusion, high-precision tests of the equivalence principle can be
viewed as windows on string-scale physics. Not only could they discover
the dilaton, but, by fitting observed data to the expected composition
dependence (\ref{eq:6.14}) of the equivalence principle violation, they
could give access to the ratios $C_B/C_E$, $C_D/C_E$ which are delicate
probes of some of the presently most obscure aspects of particle
physics: Higgs sector and unification of coupling constants.

\acknowledgments

 We wish to thank B. Julia and J. Zinn-Justin, organizers of the
July~1992 Les Houches Summer School during which this work was
initiated. T.D. thanks the CERN Theory Division for its kind
hospitality, and acknowledges helpful discussions with D.J. Gross, C.
Kounnas and G. Veneziano. A.M.P. thanks l'Institut des Hautes Etudes
Scientifiques for supporting a stimulating visit in France.
The work of A.M.P. was partially supported by the NSF Grant
PHY90211984.

\begin{figure}
\caption{The factor by which $\varphi$ is attracted
(when the early universe cools down through
 $T \sim m_A$) toward a minimum $\varphi_m$
of the function $m_A(\varphi)$ is plotted as a
 function of $b_A = \beta_A f_A^{\rm in}$.
 The solid (dashed) line corresponds to $A$ being
a fermion (boson).}
\end{figure}

\begin{figure}
\caption{ The solid
line represents $\log_{10} [ (g^2_{\rm rad} - g^2_0)/g^2_0 ]$ as a
function of $\log_{10} \kappa$, i.e.
the fractional deviation (left over at the end of the
radiation era) of the gauge coupling constants $g^2_{\rm rad} \propto
B^{-1}(\varphi_{\rm rad})$ from their present values $g^2_0$,
versus the curvature $\kappa$ of the function
$\ln B^{-1}(\varphi)$ near its minimum. The dashed line represents
an analytical estimate  (when $\kappa \protect\agt 1$)
of that deviation,
obtained by assuming that the phases $\theta$  of the WKB results
Eq.~(\protect\ref{eq:4.10a})
are randomly distributed.}
\end{figure}

\begin{figure}
\caption{The solid line represents $\log_{10}(\Delta a / a)_{\rm max}$
as a function of $\log_{10} \kappa$, i.e. the expected present level
of violation of the equivalence principle (when comparing Uranium
with a light element) as a function of the curvature $\kappa$
of the (string-loop induced) function $\ln B^{-1}(\varphi)$
near a minimum $\varphi_m$. The dashed line represents an analytical
estimate (when $\kappa \protect\agt 1$) of that violation obtained by
assuming random phases $\theta$ in Eqs.~(\protect\ref{eq:4.10a}) and
(\protect\ref{eq:5.6}). }
\end{figure}

\end{document}